\documentstyle[11pt,fleqn]{article}
\renewcommand{\thanks}[1]{\renewcommand{\thefootnote}{\fnsymbol{footnote}}
               \footnote{#1}\renewcommand{\thefootnote}{\arabic{footnote}}}
\newcommand{\preprint}[1]{\hfill{\sl preprint - #1}\par\bigskip\par\rm}
\def\titolo{\par\bigskip\begin{center}\bf\LARGE}
\def\endtitolo{\end{center}\par\bigskip\par\rm\normalsize}
\def\instit{\begin{center}\large}
\def\endinstit{\end{center}\rm\normalsize}
\def\references{\begin{thebibliography}{10}}
\def\endreferences{\end{thebibliography}\end{document}}

\newcommand{\dinfn}{\dip\\ and \infn}
\newcommand{\btit}{\begin{titolo}}
\newcommand{\etit}{\end{titolo}}

\newcommand{\Idinfn}{\begin{instit}\dinfn\end{instit}}

\renewcommand{\author}[1]{\begin{center}\Large #1\end{center}}
\renewcommand{\date}[1]{\par\bigskip\par\sl\hfill #1\par\medskip\par\rm}
\newcommand{\pacs}[1]{\smallskip\noindent{\sl PACS number(s):
                       \hspace{0.3cm}#1}\par\bigskip\rm}
\newcommand{\babs}{\hrule\par\begin{description}\item{Abstract: }\it}
\newcommand{\eabs}{\par\end{description}\hrule\par\medskip\rm}
\newcommand{\ack}[1]{\par\section*{Acknowledgments} #1}

\def\dinfn{\smallskip Dipartimento di Fisica, Universit\`a di Trento\\ 
                           and Istituto Nazionale di Fisica Nucleare,\\
                                   Gruppo Collegato di Trento, Italia}

\newcommand{\email}[1]{e-mail: \sl #1@science.unitn.it\rm}

\newcommand{\bmail}[1]{e-mail: \sl #1@fisica.uel.br\rm}

\newcommand{\sergio}{Sergio Zerbini\thanks{\email{zerbini}}}
\newcommand{\luciano}{Luciano Vanzo\thanks{\email{vanzo}}}

\newcommand{\andrei}{Andrei A. Bytsenko\thanks{\bmail{abyts}\hspace{1cm} 
On leave from  Universidade Estadual
de Londrina, Brazil}}

\renewcommand{\vec}[1]{{\bf #1}}       
\newcommand{\hs}{\qquad\qquad}         
\newcommand{\nn}{\nonumber}            
\newcommand{\beq}{\begin{eqnarray}}    
\newcommand{\eeq}{\end{eqnarray}}      
\newcommand{\beqn}{\begin{eqnarray}}   
\newcommand{\eeqn}{\end{eqnarray}}     
\newcommand{\at}{\left(}               
\newcommand{\aq}{\left[}               
\newcommand{\ct}{\right)}              
\newcommand{\cq}{\right]}              
\newcommand{\ii}{\infty}                         
\newcommand{\fr}[2]{\mbox{$\frac{#1}{#2}$}}      
\newcommand{\Tr}{\,\mbox{Tr}\,}                  
\newcommand{\PP}{\,\mbox{PP}\,}                  
\newcommand{\Res}{\,\mbox{Res}\,}                
\newcommand{\lap}{\Delta}                        

\newcommand{\be}{\beta}

\newcommand{\ep}{\varepsilon}
\newcommand{\ze}{\zeta}

\newcommand{\si}{\sigma}

\newcommand{\Ga}{\Gamma}

\newcommand{\La}{\Lambda}

\begin{document}

\preprint{UTF 342}
\btit
{\bf One-loop Quantum Holography\\
for Higher Dimensional Black Holes}  
\etit

\author{\andrei, \luciano\  and \sergio}
\Idinfn

\date{}

\babs
The one-loop quantum corrections to the free energy associated with scalar
field in a higher dimensional static curved space-time is investigated 
making use of the conformal transformation method.
For a space-time with bifurcate horizon, horizon divergences are accounted 
for choosing the Planck length as natural cutoff. The leading term in the 
high temperature quantum correction satifies holographically the  "area law",
like the tree level Bekenstein-Hawking term.
 Furthermore it 
is stressed that only for the asymptotically AdS black holes one may  
 have a  microscopic interpretation of the entropy also at quantum level.  
\eabs

\pacs{04.62.+v, 04.70.Dy}

{ \bf I}. Recently, the issue concerning the microscopic explanation of the
Bekenstein-Hawking  formula for the black hole entropy 
\cite{beke73-7-2333,hawk75-43-199} has been vastly discussed in literature.
Among the several approaches, we would like to recall the 
investigations  related to (2+1)-dimensional BTZ black hole 
\cite{carlip}, the  stringy approach
(see, for example, \cite{amanda}), the Matrix theory approach \cite{lowe}, 
the loop gravity approach
\cite{krasnov}, the induced gravity approach
\cite{jacofrolov} and the new approach appeared in \cite{mann98}.
 
There have also been some attemps to compute
semiclassically the  quantum corrections to the
Bekenstein-Hawking classical entropy for the 4-dimensional Schwarzschild
lack hole. However, so far all the
evaluations have been plagued by the appearance of  divergences, 
first noticed  in ref. \cite{thoo85-256-727} (see also
\cite{bomb86-34-373,suss94-50-2700,dowk94-11-55,furs96,empa95,cogn95}).

In the 4-dimensional black hole case, 
the physical origin of these divergences can
be traced back to the equivalence principle
according to which, in a static space-time with canonical horizons, a
system in thermal equilibrium has a local Tolman temperature  given by
$T(x)=T/{\sqrt {-{\rm g}_{00}(x)}}$, $T$ being the generic asymptotic
temperature. Since, roughly speaking, very near the horizon a static
space-time may be regarded as a Rindler-like space-time, one gets for
the Tolman temperature $T(\rho)=T/\rho$,  $\rho$ being the distance
from the horizon. As a consequence, omitting the multiplicative
constant, the total entropy reads 
\beq 
S\sim\int d{\vec x}\int_\ep^\ii
T^3(\rho)\,d\rho= \frac{AT^3}{2\ep^2} \:,\label{1.3} 
\eeq 
where $A$ is
the area of the horizon and  $\ep$ is the horizon cutoff. These
considerations suggest  the use of the optical metric
$\bar{{\rm g}}_{\mu\nu}={\rm g}_{\mu\nu}/|{\rm g}_{00}|$, conformally 
related to the
original one, 
 and one of the purposes of this paper is to implement
this idea, and try to shed some light on the  holographic property
\cite{thooholo} of black hole space-times, which has been 
discussed recently in \cite{witten,barbon} within the AdS/CFT 
correspondence \cite{malda}. \\
In particular, Barbon and Rabinovici \cite{barbon} have explained
how the internal dimensions remain invisible to the
thermodynamics of the $4$D conformal field theory living on the 
boundary
of $AdS_5$. The dimensionality of a spacetime is revealed in the free
energy of massless fields, which scales as $T^d$ in a $d$-dimensional
space. By allowing the phase transition to an anti-de Sitter black
hole, they have shown that the region where the local temperature exceedes
the Kaluza-Klein threeshold (i.e. $T>1/R_c$, where $R_c$ is the radius 
of compact
internal dimensions), is roughly inside the event horizon and thus is
cut  out from the Euclidean section, which is where the free energy
is computed. This gives one-loop holography, but only if one choses to
renormalize away the horizon divergences.\\
From the point of view of the low energy effective field theory, this
is unsatisfactory, because it requires a tree level bare entropy with 
no
statistical interpretation. Instead, we can rely on the correspondence
between ultraviolet effects in the bulk with infrared effects in the
boundary theory \cite{witten}, to argue that there should be no 
horizon divergences,
because there should be no infrared divergences in a field theory
on a compact space. Hence, we shall keep the horizon contribution and
show that it scales holographycally.

{\bf II}. To begin with, we consider a  scalar field on a $(N+1)$-dimensional 
static space-time with the metric (signature $-+..++$), $ D=N+1$. 

\beq
ds^2={\rm g}_{00}(\vec{x})(dx^0)^2+{\rm g}_{ij}(\vec{x})dx^idx^j\:, \hs
\vec{x}=\{x^j\}\:,\hs i,j=1,...,N\:. 
\eeq
The  restriction to scalar fields may be justified noting that  the horizon 
divergences we are going to discuss here should be independent on the 
specific features of the bulk (supergravity) theory.

The one-loop partition function is given by (we perform the Wick
rotation $x_0=-i\tau$, thus all differential operators one is dealing
with will be elliptic) 

\begin{equation} Z=\int d[\phi]\,
\exp\at-\frac12\int\phi L_{D} \phi d^{D}x\ct \:,
\end{equation} 
where $\phi$ is a scalar density of weight $-1/2$ and $L_{D}$ is a
Laplace-like operator that has the form

\beq L_{D}=-\lap_{D}+m^2+\xi R \:.
\eeq 
Here $\lap_{D}$ is the Laplace-Beltrami operator acting in $D$-
dimensional space-time, $m$ (the mass) and $\xi$ are arbitrary
parameters and $R$ is the scalar curvature.

We recall method of the conformal transformation 
\cite{dowk78-11-895}.
This method is useful because it permits to compute all physical
quantities in an ultrastatic manifold (called the optical manifold
\cite{gibb78-358-467}). The ultrastatic
Euclidean metric $\bar{{\rm g}}_{\mu\nu}$ is related to the static one by
the conformal transformation 

\beq
\bar{{\rm g}}_{\mu\nu}(\vec{x})=e^{2\si(\vec{x})}{\rm g}_{\mu\nu}
(\vec{x}) \:,
\eeq
with $\si(\vec{x})=-\frac{1}{2}\ln {\rm g}_{00}$. In this manner,
$\bar{{\rm g}}_{00}=1$ and $\bar{{\rm g}}_{ij}={\rm g}_{ij}/{\rm g}_{00}$ 
(Euclidean optical metric).

With regard to  the one-loop partition function, it is possible to show that 

\beq
\bar{Z}=J[{\rm g},\bar{{\rm g}}]\,Z \:,
\eeq 
where $J[{\rm g},\bar{{\rm g}}]$ is the Jacobian
of the conformal transformation. Such a Jacobian can be explicitely
computed, but here we shall need only its
structural form. Using $\zeta$-function regularization for the
determinant of the second order differential operator we get 

\beq \ln Z=\ln\bar Z-\ln J[{\rm g},\bar {\rm g}] =\frac{1}{2}
\frac{d}{ds}\ze(s|\bar L_{D}\ell^2)|_{s=0}-\ln J[{\rm g},\bar{{\rm g}}]
\:,\label{lnZ-Zbar}
\eeq 
where $\ell$ is an arbitrary parameter
necessary to adjust the dimensions, the function $\ze(s|\bar L_{D}\ell^2)$
associated with the operator $\bar L_{D}$, which explicitly reads

\beq \bar
L_{D}=e^{-\si}L_{D}e^{-\si}= -\partial_\tau^2-\bar\lap_N+\xi_{D}\bar R
+e^{-2\si}\aq m^2+(\xi-\xi_{D})R\cq =-\partial_\tau^2+\bar L_N
\:.\label{aconf}
\eeq
In Eq. (8), $\xi_{D}=(N-1)/4N$ and
\beq
\bar R=e^{-2\si}\left[R-2N\lap_{N}\si-N(N-1){\rm g}^{\mu\nu}
\partial_{\mu}\si\partial_{\nu}\si\right]\:.\label{curv}
\eeq
Now we formally have to deal with an ultrastatic space-time. 
For a scalar field
in thermal equilibrium at finite temperature $T=1/\be$ the partition function
$\bar{Z}_\be$ can be
obtained, within the path integral approach, after the  Wick rotation
$\tau=ix^0$ imposing on  the  field a  $\be$ periodicity in $\tau$. Thus 
one obtains \cite{dowk78-11-895,cogn95,byts96}
\beq 
\ln\bar
Z_\be&=&-\frac{\be}{2}\aq \PP\ze(-\fr12|\bar L_{N})
+(2-2\ln2\ell)\Res\ze(-\fr12|\bar{L}_N)\cq\nn\\ &&\hs+\lim_{s\to0}
\frac{d}{ds}\frac{\be}{\sqrt{4\pi}\Ga(s)} \sum_{n=1}^\ii\int_0^\ii
t^{s-3/2}\,e^{-n^2\be^2/4t}\, \Tr e^{-t\bar{L}_N}\,dt
\label{lnZbeta}\:, 
\eeq 
where $\PP$ and $\Res$ stand for the principal
part and for the residue of the zeta function.
The free energy is related to the canonical partition function by
means of the equation 
\beq 
F(\be)=-\frac{1}{\be}\ln
Z_\be =-\frac{1}{\be}\at\ln\bar Z_\be-\ln J[{\rm g},\bar {\rm  g}]\ct
\:.\label{FE}
\eeq 
Since we are considering a static space-time the
quantity $\ln J[{\rm g},\bar {\rm  g}]$ depends linearly 
on $\be$ and according to
Eq.~(\ref{FE}) it gives no contribution to the entropy, which  
has the form
\beq 
S_\be=\be^2
\partial_\be F_\be \:, \label{entropy} 
\eeq 
where $F_{\be}$ is the temperature dependent part (statistical sum) of 
$F(\be)$.

Let us apply this formalism to  scalar fields in
a $D$-dimensional static space-time with metric
\beq
ds^2=A(r)\,d\tau^2+
A(r)^{-1}\,dr^2+r^2\,d\Sigma_{d-1}+dE_{D-d-1} \:,\label{bh} 
\eeq
where we are using polar coordinates, $r$ being the radial one and
$d\Sigma_{d-1}$ and $dE_{D-d-1}$is are the  metrics of two 
($d-1$)-dimensional and  
($D-d-1$)-dimensional  Einstein spaces. 
Particularly interesting are  the two $D$-dimensional  space-times
relevant in
the CFT/AdS correspondence \cite{witten}: 
 $X_1=AdS_{d+1} \times M_{D-d-1}$ and
$X_2=AdS^{BH}_{d+1} \times M_{D-d-1}$. The first  contains
the  periodically
identified
AdS space  and the second one the Schwarschild  AdS black hole,   and
 $M_{D-d-1}$ is a suitable compact $(D-d-1)$-dimensional
manifold.
However, we would like to consider  a more general   class of metrics, 
which may be defined by the function $A(r)$ and by the related 
$(d-1)$-dimensional  manifold, namely 

$(I)$. The AdS space,
\beq
A(r)=\at 1+\frac{r^2}{l^2}\ct\,, \hs d\Sigma_{d-1}=d\Omega_{d-1}\,.
\label{ads}
\eeq

$(II)$. The Schwarzschild black hole,
\beq
A(r)=\at 1-\frac{C_d M}{r^{d-2}} \ct\,,  
\hs d\Sigma_{d-1}=d\Omega_{d-1}\,.
\label{S}
\eeq

$(III)$. The Schwarzschild-AdS black hole \cite{hawkpage,witten1},

\beq
A(r)=\at 1+\frac{r^2}{l^2}-\frac{C_d M}{r^{d-2}}\ct\,, 
\hs d\Sigma_{d-1}=d\Omega_{d-1}.
\label{adsS}
\eeq

$(IV)$. The toroidal AdS black hole \cite{vanzo,mann1,brill,birg},

\beq
A(r)=\at \frac{r^2}{l^2}-\frac{C_d M}{r^{d-2}}\ct\,,
\hs d\Sigma_{d-1}=dT_{d-1}.
\label{adsT}
\eeq

$(V)$. The hyperbolic AdS black hole \cite{vanzo,mann1,brill,birg},

\beq
A(r)=\at-1+ \frac{r^2}{l^2}-\frac{C_d M-\frac{2}{d}\at(\frac{d-2}{d})
l^2 \ct^{\fr{d-2}{2}}}{r^{d-2}}\ct\,,
\hs d\Sigma_{d-1}=dH_{d-1}.
\label{adsTop}
\eeq
In Eqs. (15) - (18) $M$ is the mass of the black hole, the cosmological 
constant 
$\La$ is given by $|\La|=1/l^2$, $C_d$ is a normalization constant
depending on the $(d+1)-$ dimensional Newton Constant and $d\Omega_{d-1}$,
$dT_{d-1}$ and $dH_{d-1}$ are the the metric of the (d-1)-dimensional sphere,
 torus and compact hyperbolic manifold respectively.

Generally speaking, the optical metric reads

\beq
d\bar s^2=d\tau^2+\frac{dr^2}{A(r)^2}+\frac{r^2}{A(r)} d\Sigma_{d-1}
+\frac{1}{A(r)} dE_{D-d-1}
 \,.
\eeq
When a metric has an event horizon and the black hole is not extremal, 
the localization of the horizon is
given by the simple positive root of $g^{11}$, namely $A(r_+)=0$. Thus one
can make use of the near-horizon  approximation, which is valid  for 
large black hole mass. As a result
the optical metric may be approximated by (see, for example, Refs. 
\cite{cogn95,byts96})

\beq
d\bar s^2 \simeq d\theta^2+\frac{d\rho^2}{\rho^2}+\frac{r_+^2}{\rho^2}
d\Sigma_{d-1}+\frac{1}{\rho^2} dE_{D-d-1}
\,,
\eeq
where

\beq
\rho=2(r-r_+)^{1/2}\left[\frac{d}{dr}A(r)|_{r=r_+}\right]^{-1/2}\,, 
\hs \theta=\frac{\tau}{2}\frac{d}{dr}A(r)|_{r=r_+}\,.
\eeq
In this not extremal case, one gets in a standard way the 
inverse of the Hawking temperature requiring the
absence of the conical singularity, namely 
$\be_H=4\pi \at dA(r)/dr|_{r=r_+} \ct^{-1}$.

In order to study the quantum properties of matter fields, it is 
sufficient to 
investigate the kernel
of the operator $e^{-t\bar{L}_N}$ and use the Eq. (\ref{lnZbeta}). We shall  
to assume the high temperature expansion or, in the case of 
black hole, the Rindler-like approximation. 

In both the circumstances, the leading term of the heat-kernel turns to be 
\cite{dowk78-11-895,cogn95}

\beq
\Tr e^{-t\bar L_N}\simeq \frac{\bar V_{N}}{(4\pi t)^{N/2}\,,}
\eeq
where $\bar V_N$ is the optical volume of the whole spatial section. 
From Eq. (10) the corresponding
off-shell free energy reads

\beq
\bar F_\be \simeq-\frac{\bar V_N}{\be^{D}}\,.
\eeq
For a space-time without event horizon (the AdS space) there are no horizon
divergences and the free energy has a leading term $\be^{-D}$, i.e. no
holographic reduction of the demensionality is present.

The situation is different for  black hole space-times. 
The optical volume, formally given by

\beq
\bar V_N =V_{d-1} V_{D-d-1}\int_{r_+}^\ii dr \frac{r^{d-1}}{A(r)^{D/2}}\,,
\eeq
is divergent, because of the non integrable singularity  at $r=r_+$. 
Thus one has to introduce a cutoff parameter, 
i.e.  $r_+ = \ep$. In fact, 
making use of the near-horizon  approximation, it is possible to show that 
the leading term in the off-shell free energy is \cite{byts96}

\beq
\bar F_\be \simeq -\frac{r_+^{d-1} V_{d-1}V_{D-d-1}}{\ep^{D-2}}
\frac{\be_H^{D-1}}{\be^{D}}\,.
\eeq 
The off-shell entropy can be computed by means of Eq. (\ref{entropy}) and 
the result is
\beq
S_\be \simeq \frac{r_+^{d-1}V_{d-1}V_{D-d-1}}{\ep^{D-2}}
\left(\frac{\be_H}{\be}\right)^D\,.
\eeq 
From a phenomelogical point of view, well known stringy arguments suggest 
to choose $\ep\simeq l_P$ (of the order of the  $D$-dimensional 
Planck length) \cite{thoo85-256-727}. Furthermore,   
$G_{D} \simeq l_P^{D-2}$ and, and recalling the relation between the 
Newton constants  in different dimensions
\beq
\frac{V_{D-d-1}}{G_D}=\frac{1}{G_{d+1}}
\eeq
and finally going on-shell at $\be=\be_H$,
  one obtains the 
"area law-like" expression also for the one-loop quantum 
correction:

\beq
S \simeq \frac{r_+^{d-1}V_{d-1}}{G_{d+1}}\,,
\eeq
where the proportionality factor depends on species of fields. 
As a consequence, going on-shell, one obtains  the holographic reduction of
 the dimensionality,  in agreement
with the  results of ref. \cite{barbon}, even though the dimensionality 
reduction mechanism is completely different. In fact, in our approach, 
the dimensionality reduction is due to the presence of quantum
 horizon divergences.

$\bf III$. We conclude with some remarks. First, as far as the computation 
of the quantum black hole entropy is concerned, the "internal" 
$D-d-1$-dimensional  manifold does not contribute. Second, 
the class of D-dimensional blac k
hole we have considered and for which we have estimated the one-loop quantum
correction to the entropy, can be divided into two subclasses: the first one, 
example 
$(II)$ (Schwarzschild black hole), corresponds to non-negative cosmological 
constant and the second one, examples $(I), (III)-(V)$ (toroidal,
Schwarzschild-AdS and hyperbolic AdS black holes), corresponds to a negative 
cosmological constant. For all solutions, one can easily computed a 
relationship between the $r_+$ and the Hawking temperature, eliminating the
mass of the black hole. This is not a trivial task, particularly for the 
hyperbolic AdS black holes (see \cite{vanzo} and Eq. \ref{adsTop}). 
Omitting the details,
it turns out that for the first type of black holes 
$r_+ \simeq \be_H$, and for the second type $r_+\simeq {\be_H}^{-1}$.
As a consequence, the entropy for the Schwarzschild black hole goes like
\beq
S\simeq T^{-d+1}\,,
\eeq
related to negative specific heat and the instability of this  black hole. 
With regard to  the asymptotically AdS 
black holes,
\beq
S\simeq T^{d-1}\,,
\eeq
which , in turn, is related to the positivity of the their specific heat 
and their stability. 

As a consequence, only for the stable AdS black hole it seems to exist, 
also  at 
quantum level, a microscopic
explanation of the black hole entropy via the AdS/CFT correspondence. In fact
 (conformal) quantum fields 
living on the horizon (boundary) of the black hole have a statistical entropy
with $S\simeq T^{d-1}$ as leading term.

\ack{We would like to thank G. Cognola, D. Fursaev  and M. Tonin
for discussions. A.A. Bytsenko wishes to thank CNPq (Brazil) and INFN (Italy)
for financial support. The research of A.A. Bytsenko was supported in part
by RFFI (grant No. 98-02-18380-a) and by GRACENAS (grant No. 6-18-1997).}

\end{document}